\documentstyle[12pt,epsf,psfig]{article}
\newcommand{\be}{\begin{equation}}
\newcommand{\ee}{\end{equation}}    

\begin{document}
 
\begin{center}
{\Large\bf An inhomogeneous and anisotropic Jastrow function
for non-uniform many-electron systems} \\

\vspace{0.5cm}
{ A. Garcia-Lekue,$^{\rm (1)}$
M. Nekovee,$^{\rm (2)}$
J.M. Pitarke,$^{\rm (1,3)}$
and
R. Gaudoin$^{\rm (4)}$} \\
                                            
\vspace{0.6cm}
{\it $^{\rm (1)}$ Materia Kondentsatuaren Fisika Saila, Zientzi Fakultatea,
Euskal Herriko Unibertsitatea, 644 Posta kutxatila,\\
48080 Bilbo, Basque Country, Spain}\\
{\it $^{\rm (2)}$ Centre for Computational Science, Department of Chemistry,
Queen Mary and Westfield College, London, UK 
}\\
{\it $^{\rm (3)}$ Donostia International Physics Center (DIPC) and Centro Mixto
CSIC-UPV/EHU, Donostia, Basque Country, Spain 
}\\
{\it $^{\rm (4)}$  CMTH Group,
Blackett Laboratory, Imperial College, London, UK 
}\\
\end{center}
                                   
\begin{abstract}
Quantum Monte Carlo simulations of interacting electrons in solids
often use Slater-Jastrow trial wave functions. The Jastrow function
takes into account correlations between pairs of electrons.
In simulations of solids, it is common to use a Jastrow function which is
both homogeneous and isotropic.
The use of a homogeneous and isotropic Jastrow factor is questionable
in applications to systems with strong density inhomogeneities, such
as surfaces and multilayers.
By generalizing the original derivation of the RPA Jastrow factor
for the homogeneous electron gas \cite{Gaskell} 
to inhomogeneous systems we
derive an scheme for generating inhomogeneous and anisotropic Jastrow
factors for use in nonuniform systems, from the non-interacting
static structure factor and density of the system.
We discuss aspects of our scheme and illustrate it with a first
application to an inhomogeneous electron gas.

\end{abstract}

\section{Introduction}

In this paper we consider Slater-Jastrow trial wave functions, 
$\Psi= D e^J$, where $D$ is a Slater determinant and $J$ the Jastrow
factor that takes account of the electronic correlations. Such wave functions
are employed in variational quantum Monte Carlo (VQMC) simulations
of solids \cite{VQMC1},\cite{VQMC2}, where an approximate many-electron
ground state is obtained by numerically optimizing an explicit parametrized
trial wave function. Remarkably better results are obtained from VQMC
calculations using Slater-Jastrow functions than from Hartree-Fock (HF)
or density-functional theory within the local density approximation (LDA). 
For example, 
cohesive energies obtained using 
VQMC are typically an order of magnitude more accurate than  those
obtained employing  HF or LDA \cite{cohesive1}-\cite{cohesive5}.
 
We use a Slater determinant of LDA orbitals and a Jastrow
factor that includes pairwise correlation terms $u({\bf r_i},{\bf r_j})$.
In simulation of solids $u$ is usually taken to be both homogeneous and 
isotropic, i.e., $u({\bf r_i},{\bf r_j})$ is assumed to depend only on the
interelectronic distance $r_{ij}=|{\bf r_i}-{\bf r_j}|$.

The long-range behaviour of $u({\bf r_i},{\bf r_j})$ 
is determined by the zero-point motion of the plasmons, both 
in homogeneous \cite{Bohn-Pines} and inhomogeneous systems
\cite{rene} while the short-range behaviour 
is governed by the so-called cusp condition \cite{cusp}.
Gaskell  minimized the variational energy of the uniform 
electron gas in the random-phase approximation and
found a Jastrow function which combined both the short-range and 
the long-range behaviour. Ceperley used this Jastrow function
in his variational Monte Carlo calculations of the 
spin-unpolarized homogeneous electron gas \cite{Ceperley}.

Here we generalize Gaskell's approach to inhomogeneous systems and 
derive an inhomogeneous and anisotropic Jastrow factor that depends on the
non-interacting structure factor and density of the system.
The Jastrow factor is represented as a double Fourier integral. 
The full interacting many-electron Hamiltonian $\hat H$ is
expressed as a function of the plane-wave expansion coefficients of the 
Jastrow function 
and, within the random phase approximation (RPA) $\hat H$ is minimized
with respect to these coefficients.
This way the approximated ground state  energy is obtained, 
and the trial  wavefunction of the system can be constructed from the 
resulting inhomogeneous and anisotropic Jastrow function.
In the homogeneous limit, our approach reduces to that of Gaskell.

The rest of the paper is organized as follows. In Sec.2.1. we describe
the formalism used  in Gaskell's original derivation  of 
the Jastrow function for the homogeneous electron gas (HEG).
Sec. 2.2 presents our generalization to systems with density 
inhomogeneities where, for the sake of simplicity, we consider 
systems with density variations in only one direction.
Sec. 3 discusses the numerical results obtained for the IHEG, and  
Sec.4 concludes. 
\section{Formalism}

A Slater-Jastrow trial wave function is the product of a totally antisymmetric
Slater determinant $D$ and a totally  symmetric Jastrow factor $e^J$:
 $\Psi=D e^J$.
The orbitals used in D are obtained from LDA  calculations. Here
we consider spin unpolarized systems.
The Slater determinant builds in exchange effects but neglect the electronic
correlations caused by the Coulomb interactions. The most important correlation
effects occur when pairs of electrons approach each other, and these may be
included by choosing pairwise Jastrow factors of the form:
\be\label{eq2}
 J=-\sum_{i,j} u({\bf r}_{i},{\bf r}_{j}), 
\ee
where $ u({\bf r}_{i},{\bf r}_{j})= u({\bf r}_{j},{\bf r}_{i})$.
So the trial wave function can be written as:
\be\label{eq3}
 \Psi({\bf R})= D \exp\left[-\sum_{i,j} u({\bf r}_{i},{\bf r}_{j})
\right] \;.    
\ee 
Note that we include the $i=j$ terms in the sum over $i$ and $j$, which 
are omitted by many authors as they give a constant contribution for the
HEG.
(Unless otherwise is stated, we use atomic units throughout, i.e.,
$e^2=\hbar=m_e=1$.)

The full interacting many-electron  Hamiltonian is given by
\be\label{eq4}
 \hat H= \hat T + \hat V_{ee} + \hat V_{ext},
\ee  
where $\hat T$ and $\hat V_{ee}$ are the kinetic energy and the 
electron-electron interaction operators respectively and 
$\hat V_{ext}$ is an external potential.  The expectation value
of the energy with respect to the above trial wavefunction can be
written as
 (the normalized $\Psi$ is taken):
\be\label{eq5}
E=<\Psi|H|\Psi>=\sum_{i=1}^{N} \epsilon_{i} + \int_{}^{}d{\bf r}n({\bf r})
    V_{ext}({\bf r}) - \int_{}^{}d{\bf r}n({\bf r})V_{s}({\bf r}) 
     + <T_2> + <V_{ee}> ,
\ee          
where $V_{s}$ is the effective single-particle potential associated with 
the orbitals in the Slater
determinant, $\epsilon_{i}$ are the energy
eigenvalues which correspond to such one electron orbitals and 
$n({\bf r})$ is the electron density. $<T_{2}>$ and $<V_{ee}>$ are defined as:
\be\label{eq6}
 <T_{2}>= \int_{}^{}d{\ R} \Psi^{\ast}({\bf R})\Psi({\bf R})\left[
  \sum_{m=1}^{N}{\bf \nabla}_{m}(\sum_{i,j}U({\bf r}_{i},{\bf r}_{j}))\cdot
     {\bf \nabla}_{m}(\sum_{i,j}U({\bf r}_{i},{\bf r}_{j}))\right]\\
\ee
and
\be
 <V_{ee}>=  \int_{}^{}d{\bf R} \Psi^{\ast}({\bf R})\Psi({\bf R})
    (\frac{1}{2}\sum_{i\neq j}^{}\frac{1}{|{\bf r}_{i}-{\bf r}_{j}|}).
\ee
Here ${\bf R}= ({\bf r_1},{\bf r_2}, \ldots {\bf r_N})$
represents all electron coordinates.
Several different types of Jastrow factors are commonly used, 
the best known are based on the RPA  in the form due to Gaskell and
the form based on the RPA of Bohm and Pines \cite{Bohn-Pines}.
Here we will briefly describe the Jastrow factor due to Gaskell for 
the homogeneous electron gas  
before presenting its generalization to inhomogeneous systems.

\subsection{The homogeneous RPA Jastrow factor}
For  homogeneous systems the correlation term $u$ depends only on the 
interelectronic distance $r_{ij}$ and  may be expressed as follows:
\be\label{eq7} 
u({\bf r_i},{\bf r_j})=u(r_{ij})=\sum_{k}u(k)e^{{\rm i}kr_{ij}}, 
\ee 
where $k$ is the modulus of the total wavevector. Also, 
the single-particle orbitals are a set of planewaves. Inserting 
the above expression in Eq.(\ref{eq5}), the total energy 
in the random-phase approximation of $T_2$ can be written as
\cite{Gaskell}:
\be\label{eq8}
E= \sum_{k\leq k_F}\frac{1}{2}k^2+2N^2\sum_{{\bf k}}k^2 u^2(k) S(k)+
 2\pi n_{0}\sum_{{\bf k}}\frac{1}{k^2}[S(k)-1]
 \ee
where $NS(k)=<\Psi|\hat n_{\bf k}\hat n_{\bf k}^*|\Psi>$ 
is the static structure factor of the uniform interacting electron gas
and $\hat n_{\bf k}=\sum_{i}e^{{\rm i}{\bf k}\cdot{\bf r}_i}$ are the number
density operators. $k_F$ and $n_0=N/V$ are  the Fermi energy and the 
average electron densityi, respectively, $N$ being the total number 
of electrons in  the system and $V$ the volume.
Gaskell assumed that the structure factor for the interacting electron
gas is close to that of the non-interacting electron gas and, for a small
Jastrow function $u({\bf r})$
 \footnote{This corresponds to the high density limit.}  
can be approximated by the perturbation formula 
\cite{Gaskell}:
\be
 S(k)=\frac{S_0(k)}{1+4u(k)S_0(k)},
\ee
where $S_0(k)$ is the structure factor of the non-interacting electron
gas \cite{Pines-Nozieres}.
The variational energy in the random-phase approximation (Eq. \ref{eq4})
can then be minimized with respect to 
$u(k)$ to yield:
\be\label{ukmin}
 u(k)=\frac{-1}{4S_0(k)}+\frac{1}{4}\left[\frac{1}{4(S_0(k))^2}+
 \frac{16\pi  n_0}{k^4}\right]^{1/2}.
\ee
\subsection{The inhomogeneous RPA Jastrow factor}

Here we consider
the case   
of an inhomogenous system  translationally invariant 
in the plane perpendicular to the $z$ axis. In this case 
$u$ can be written as:
\begin{equation}\label{Fourier}
u({\bf r_1},{\bf r_2}) = \sum_{{\bf q}}\sum_{k_{z_{1}},k_{z_{2}}}u_{k_{z_{1}},k_{z_{2}}}
e^{{\rm i}{\bf q}\cdot({\bf r_\parallel 1}-{\bf r_{\parallel 2}})}e^{{\rm i}k_{z_{1}}z_1}e^{{\rm i}k_{z_{2}}z_2},
\end{equation}
where ${\bf r}=({\bf r_\parallel},z)$ and ${\bf k}=({\bf q},k_{z})$.
$u_{k_{z_{1}},k_{z_{2}}}$ are the two-dimensional Fourier coefficients
of the correlation function $u$:
\be
 u_{k_{z_{1}},k_{z_{2}}}(q)=\frac{1}{V}\int
 d{\bf r_\parallel}e^{{\rm i}{\bf q}
 \cdot{\bf r_\parallel}}\int dz_1\int dz_2
         e^{{\rm i}k_{z_1}z_1}e^{-{\rm i}k_{z_2}z_2}
u({\bf r_\parallel},z_1,z_2).
\ee  
We use the following definition of the number density operator:
$\hat{n}_{{\bf q},k_{z}} = \sum_{i}e^{i {\bf q} \cdot {\bf r_\parallel}_i }
 e^{i k_z z_i}$,
so that the Slater-Jastrow trial wave function is given by:
\begin{equation}
\Psi = D {\rm exp}\left[-\sum_{\bf q} \sum_{k_{z_{1}},k_{z_{2}}}
  u_{k_{z_{1}},k_{z_{2}}}({\bf q})
\hat{n}_{{\bf q},k_{z_1}} \hat{n}^*_{{\bf q}k_{z_2}}\right].
\end{equation}

\subsubsection{Deriving an approximation for the inhomogeneous
structure factor}
In this section we develop an approximation based on the RPA to obtain 
the interacting structure factor in terms of Fourier coefficients of 
the above Jastrow factor.
Let us  define:
\begin{equation}
\Psi(\lambda) = D {\rm exp}\left[-\lambda\sum_{\bf q} \sum_{k_{z_{1}},k_{z_{2}}} 
 u_{k_{z_{1}},k_{z_{2}}}({\bf q})
\hat{n}_{{\bf q},k_{z_1}} \hat{n}^*_{{\bf q}k_{z_2}}\right].
\end{equation}
The expectation value we need to calculate is of the form:

\begin{equation}
I(\lambda) \equiv \frac{< \Psi(\lambda)|\hat{n}_{{\bf q},k_{z_1}} 
\hat{n}^*_{{\bf q}k_{z_2}}|
 \Psi(\lambda)>}{<\Psi(\lambda)|\Psi(\lambda)>}.
\end{equation}
with $\lambda=1$. With the above trial function this becomes 
\begin{equation}
I(\lambda)=
\frac{\int_{}^{} DD^*\hat{n}_{{\bf q_1},k_{z_1}} \hat{n}^*_{{\bf q_2}k_{z_2}}
 {\rm exp}\left[-2\lambda\sum_{{\bf q_2}} \sum_{{k_{z_3},k_{z_4}}}
  u_{k_{z_3},k_{z_4}}({
\bf q_2})
\hat{n}_{{\bf q_2},k_{z_3}} \hat{n}^*_{{\bf q_2}k_{z_4}}\right]}
 {\int_{}^{}DD^*
 {\rm exp}\left[-2\lambda\sum_{{\bf q_2}} \sum_{{k_{z_3},k_{z_4}}}
 u_{k_{z_3},k_{z_4}}({
\bf q_2})
\hat{n}_{{\bf q_2},k_{z_3}} \hat{n}^*_{{\bf q_2}k_{z_4}}\right]}.
\end{equation}  
  
Following Gaskell we use the identity:
\be\label{identity}
 \frac{1}{I(\lambda)}-\frac{1}{I(0)}=\int_{0}^{\lambda}\frac{\partial}
 {\partial \lambda'}\left[\frac{1}{I(\lambda ')}\right] d\lambda '
\ee                        
to obtain a perturbative approximation for $I(1)$. Taking the
derivative inside the integral with respect to $\lambda$ and 
expanding in a Taylor series the leading term is
\be
\begin{array}{ll}
\frac{\partial}{\partial\lambda'}
(\frac{1}{I(\lambda')})|_{\lambda=0}=&
4 u_{k_{z_1},k_{z_2}}({\bf q_1})
\left[\frac{<0|(\hat{n}_{{\bf q_1},k_{z_1}} \hat{n}^*_{{\bf q_1}
k_{z_2}})^{2}|0>-
<0|\hat{n}_{{\bf q_1},k_{z_1}} \hat{n}^*_{{\bf q_1}k_{z_2}}|0>^{2}}
{<0|\hat{n}_{{\bf q_1},k_{z_1}} \hat{n}^*_{{\bf q_1}k_{z_2}}|0>^{2}}\right]\\
&\\
    & + 2 \sum_{
     {\bf q_1},k_{z_3},k_{z_4}
     ({\bf q_2} \neq \pm {\bf q_1},
      k_{z_3}\neq k_{z_1},
      k_{z_4}\neq k_{z_2})}
      u_{k_{z_3},k_{z_4}}({\bf q_2}) \\
& \\
& \times \left[\frac{<0|\hat{n}_{{\bf q_2},k_{z_3}} \hat{n}^*_{{\bf q_2}
k_{z_4}}
  \hat{n}_{{\bf q_1},k_{z_1}} \hat{n}^*_{{\bf q_1}k_{z_2}}|0>}
{<0|\hat{n}_{{\bf q_1},
k_{z_1}} \hat{n}^*_{{\bf q_1}k_{z_2}}|0>^{2}}-
 \frac{<0|\hat{n}_{{\bf q_2},k_{z_3}} 
\hat{n}^*_{{\bf q_2}k_{z_4}}|0>
<0|(\hat{n}_{{\bf q_1},k_{z_1}} 
\hat{n}^*_{{\bf q_1}k_{z_2}}|0>)}
{<0|\hat{n}_{{\bf q_1},k_{z_1}}      
 \hat{n}^*_{{\bf q_1}k_{z_2}}|0>^{2}}\right]
\end{array}
\ee 
where $<0|\hat{A}|0>$ denotes the expectation value of an operator 
$\hat{A}$ with respect to $\Psi(0)=D$. 
Following Gaskell, we use the following approximation:
\be
 <0|(\hat{n}_{{\bf q},k_{z_1}} \hat{n}^*_{{\bf
q}k_{z_2}})^{2}|0>=2\left[<0|(\hat{n}_{{\bf q},k_{z_1}} \hat{n}^*_{{\bf
q}k_{z_2}})|0>\right]^{2}+ \bigcirc(1/N),
\ee
and ignore correlations between different $\hat{n}'s$ in the second 
term. 
The resulting approximation for $I(\lambda)$ becomes  
\be
 I(\lambda)=\frac{I(0)}{1+\lambda I(0) 4 u_{k_{z_1},k_{z_2}}({\bf q})}.
\ee       
 
The corresponding approximation for the Fourier coeffecients of 
the structure factor is obtained form the above equation for
$\lambda=1$, yielding 
 
\be\label{sqqiheg}
 S_{k_{z_{1}},k_{z_{2}}}({\bf q})=\frac{ S^{0}_{k_{z_{1}},k_{z_{2}}}({\bf q})}{ 1+ 4 u_{k_{z_{1}},k_{z_{2}}}({\bf q})S^{0}_{k_{z_{1}},k_{z_{2}}}({\bf q})}.
\ee

\subsubsection{Expectation value of the energy}

 In order to obtain the expression for the total energy of an inhomogeneous
 system we use  the RPA approximation
\be
 <\hat {n}_{{\bf q}_{1},k_{z_{1}}}\hat{n}_{{\bf q}_{2},k_{z_{2}}}\hat{n}_{{\bf q}_{1}+{\bf q}_{2},k_{z_{3}}}>
=<\hat{n}_{{\bf q}_{1},k_{z_{1}}}\hat{n}_{{\bf q}_{2},k_{z_{2}}}>
    <\hat{n}_{{\bf q}_{1}+{\bf q}_{2},k_{z_{3}}}>.
\ee
In the case of a system with density variation along $z$ we have:
\be
 <\hat{n}_{{\bf q}_{1}+{\bf q}_{2},k_{z}}> =\left\{ \begin{array}{ll}
                V\rho_{k_{z}} & \mbox{ if ${\bf q}_{1}+{\bf q}_{2}=0$}\\
                0 & \mbox {otherwise}
                \end{array}
                \right.
\ee                   
where $\rho_{k_{z}}$ represent the Fourier coefficients of the electronic
density.   

Making these assumptions and employing the expression for the structure
factor obtained in the previous section, the terms $<T_{2}>$ and $<V_{ee}>$
which appear in the expression for the total energy Eq.(\ref{eq5})
can be written as:
\be
 \begin{array}{rl}
 <T_{2}>=&
  2 V N \sum_{{\bf q}_{1},k_{z_{1}},k_{z_{2}},k_{z_{3}},k_{z{4}}}^{}
  \hat u_{k_{z_{1}},k_{z_{2}}}(q_{1})\hat u_{k_{z_{3}},k_{z{4}}}(q_{1})
   \left[(q_{1})^{2}-k_{z_{1}}k_{z_{3}}\right] \\
& \\
  &\times \rho_{k_{z_{1}}+k_{z_{3}}}
  \left (S_{k_{z_{2}},-k_{z_{4}}}(q_{1})\right )\\
&\\
<V_{ee}>=&n_{0}\sum_{{\bf q},k_{z}}\frac{2\pi}{(q^{2}+k_{z}^{2})^{2}
}
 \left[ \left (S_{k_{z},-k_{z}}(q) \right )-1\right].
\end{array}    
\ee
The above equation combined with our approximation (Eq.\ref{sqqiheg}) for 
the structure factor leads to an expression for the total energy 
in terms of 
 the Fourier coefficients of the $u$ function,
the electron density and the non-interacting structure factor.
We take both the density and the non-interacting structure factor from
LDA calculations and minimize  $E$ with respect to the $\hat u_{k_{z},k_{z'}} 
(q)$ coefficients, which then can be used to construct the trial 
wave functions for VQMC calculations.

\section{Results}
This section presents the numerical results obtained for uniform and 
non-uniform spin-unpolarized $N-$electron systems ($N=38$)
contained in a simple cubic (SC) simulation cell and satisfying periodic boundary 
conditions within the cell. 
Density modulations  in $z$ are induced by subjecting the system 
to and external potential of the from 
\begin{equation}
 V_{ext}({\bf r})=V_0 cos(qz)
\end{equation}
with $V_0=0.1$ a.u.. $q =0.6 k_F$ is a reciprocal lattice vector of 
the SC simulation cell. The average density of the system corresponds 
to $r_s=2$. The electron density is uniform in the $x-y$ plane and 
varies in th $z$ direction.

The inhomogeneous $u({\bf r_1},{\bf r_2})$ term is represented by fixing
the position ${\bf r_1}$ of the first  electron, while moving the
second electron, ${\bf r_2}$, along the $z$ direction on a line
passing through ${\bf r_1}$.

\subsection{Homogeneous system}
   
\begin{figure}[hbt]
\centerline{\psfig{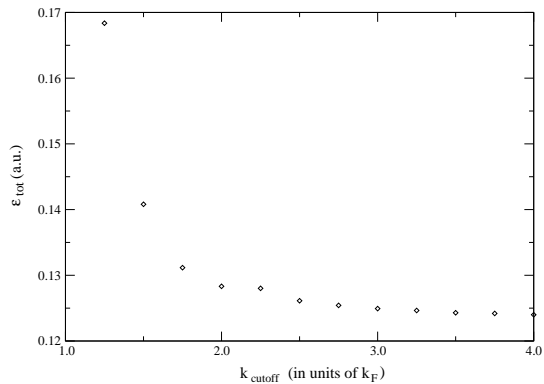}}
\caption{\protect\small
The convergence of the total energy as a function of the
cutoff $k_{cutoff}$ used in the truncated Fourier series representation
of the $u$ function. The results are for $N=38$ and $r_s=2$. $k_{cutoff}$
is given in units of $k_F$.} 
\label{fig:fig1}
\end{figure}

In Eq.(\ref{eq7}) we saw that $u(z_1,z_2)$  may be expanded as a Fourier
series. In Fig. \ref{fig:fig1} the convergence of the energy with 
respect to the cutoff wavevector, $k_{cutoff}$, used in the 
Fourier expansion of the $u$ factor is shown. The typical cutoff 
needed is small enough to be computationally feasible, so that
such a representation for $u(z_1,z_2)$  is found to be very useful. 

Looking at the expression for the energy we see that for 
$u(k)=0$, $<T_2>=0$ and $<V_{ee}>$ is just the exchange energy of a HEG.
For non-zero $u(k)$ factors \mbox{$<T_2>+<V_{ee}>$} includes both exchange and
correlation energies. The correlation energy is  defined as:
\be
E_c=[<T_2>+<V_{ee}>]-[<T_2>+<V_{ee}>]_{u(k)=0}
\ee                
Table 1 presents the values for the correlation energy obtained in 
this work in  comparison  with those obtained by Gaskell\cite{Gaskell} .
Differences due to finite size effects have to be taken into account, as
we have considered a very small system with only $N=38$ electrons.

\begin{table}
\begin{center}
\begin{tabular}{||c|c|c||}      \hline
$\ \ r_s  \ \ $&$\ \ \epsilon_{c}$ \ \ & $\ \ \epsilon_{c}(Gaskell)\ \ $\\ \hline
2&-0.04677&-0.058\\ \hline
3&-0.04175&-0.05\\ \hline
4&-0.03809&-0.0445\\ \hline
5&-0.0352&-0.0405\\ \hline
\end{tabular}
\end{center}
\caption{\protect\small
Correlation energy per electron in a.u.}
\end{table}

The results for the energy obtained with the uniform system previously
described and using the homogeneous RPA u factor given in Section. 2.1. 
are presented in Table 2.

\subsection{Inhomogeneous systems}

The inhomogeneous RPA jastrow factor considered here is the one derived
in Section. 2.2. The energies derived are presented in Table 2.
Exchange-correlation energy is defined as:
\be
E_{xc}=<T_2>+<V_{ee}>-E_H, 
\ee
where $E_H$ represents  Hartree energy.
\begin{table}[hbt]
\begin{center}
\begin{tabular}{||c|c|c|c|c||}      \hline
&$\epsilon
_{kin}^{0}$&$\epsilon_{xc}$&$\epsilon_{c}$&$\epsilon_{tot}$\\
 \hline
I)&0.2807&-0.157&-0.04677&0.1235\\ \cline{2-5}
II)&0.2658&-0.1193&-0.0061&0.144\\ \hline
\end{tabular}
\end{center}     
\caption{\protect\small
Non-interacting kinetic energy $\epsilon_{kin}^{0}=\frac{1}{N} \sum_{k}k^2/2$,
  exchange-correlation
energy $\epsilon_{xc}=E_{xc}/N$, correlation energy $\epsilon_{c}=E_c/N$
and total energy $\epsilon_{tot}=E/N$   
per electron in a.u. for I) an uniform system using
the homogeneous RPA $u$ factor, II) a weakly inhomogeneous system
using  the inhomogeneous RPA $u$ factor. The cutoff wavevector used 
was equal to $2.5k_F$. }
\end{table}

\begin{figure}
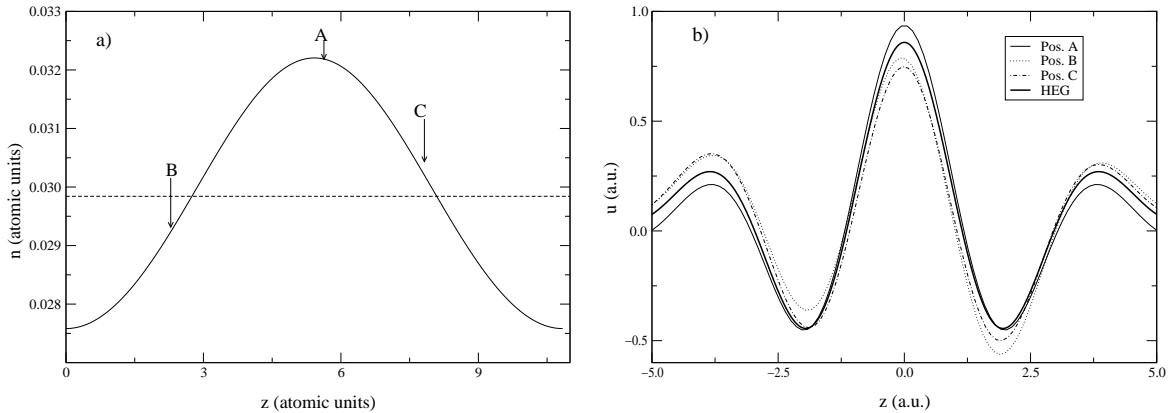

\begin{tabular}{cc}
\psfig{figure={fig2.eps},width=7.5cm,angle=270} & 
\psfig{figure={fig3.eps},width=7.5cm,angle=270}\\
\end{tabular}
\caption{\protect\small
a) LDA electronic density of the system. The horizontal
line represents the average electronic density of the system.
b) Inhomogeneous RPA u term for the three different positions
of the fixed electrons shown in a), together with the homogeneous result
for the
same system.} 
\label{fig:fig23}   

\end{figure}

In Fig.\ref{fig:fig23}a we  plot the inhomogeneous RPA $u$ term, for
the three different positions of ${\bf r_1}$ indicated in Fig. 
\ref{fig:fig23}b 
 which presents the LDA density and the average electron density.   
The homogeneous Jastrow factor is shown as well, which 
depends only on the relative distance between the electrons and 
not their individual coordinates.
Anisotropy of the Jastrow factor can 
be observed. This is 
such that the $u$ function is stronger on the side where the density
is lower. This can be explained by the fact that the RPA screening
is more effective where the electron density is high, so that the $u$
function is weaker on the high density side \cite{rene}.

\section{Conclusions}

We generalize the RPA treatment of the homogeneous electron gas by
Gaskell to the inhomogeneous case. As a result we obtain a Slater-Jastrow
trial wave functions containing  an anisotropic and inhomogeneous 
Jastrow factor. The latter may be obtained from the non-interacting
structure factor and density of the system.

Results for the energy and the inhomogeneous correlation
$u$ factor using an electronic system  subject  to a  sinusoidal
external potential are presented. We compare these results with those obtained
using the homogeneous $u$ factor by Gaskell, and we clearly observe
the  anisotropy introduced by the inhomogeneous
$u$ factor. 
Future work will focus on testing the performance of our inhomogeneous Jastrow 
factor using variational Monte Carlo and in further 
refining our approximation of the interacting structure factor.

\section{Acknowledgements} 
We acknowledge partial support by the Basque Unibertsitate eta
Ikerketa Saila.


\begin{thebibliography}{99}
\setlength{\baselineskip}{0.4cm}
\bibitem{Gaskell} T. Gaskell, Proc. Phys. Soc. {\bf 77}, 1182 (1961); Proc.
Phys. Soc. {\bf 80}, 1091 (1962).
\bibitem{VQMC1} M.H. Kalos and P.A. Whitlock, \emph{Monte Carlo Methods
 Volume 1: Basics} (Wiley, New York, 1986).
\bibitem{VQMC2} B.L. Hammond, W.A. Lester, Jr., and P.J. Reynolds,
{\emph Monte Carlo Methods in Ab Initio Quantum Chemistry} (World
Scientific, Singapore, 1994).
\bibitem{cohesive1} S. Fahy, X.W. Wang, and S.G. Louie, Phys. Rev. Lett. 
{\bf 61}, 1631 (1988).
\bibitem{cohesive2} S. Fahy, X.W. Wang, and S.G. Louie, Phys. Rev. B
{\bf 42}, 3503 (1990).
\bibitem{cohesive3} X.P. Li, D.M. Ceperley, and R.M. Martin, Phys. Rev. B
{\bf 44}, 10929 (1191).
\bibitem{cohesive4} G. Rajagopal, R.J. Needs, A. James, S.D. Kenny, and
W.M.C. Foulkes, Phys. Rev. B {\bf 51}, 10591 (1995).
\bibitem{cohesive5} P.R.C. Kent, R.Q. Hood, A.J. Williamson, R.J. Needs,
W.M.C. Foulkes, and  G. Rajagopal, Phys. Rev. B {\bf 59}, 1917 (1999). 
\bibitem{Bohn-Pines} D. Bohm and D. Pines, Phys. Rev. {\bf 92}, 609
(1953); A. Malatesta. S. Fahy, and G.B. Bachelet, Phys. Rev. B 
{\bf 56}, 12201 (1997).
\bibitem{rene} R. Gaudoin, M. Nekovee, W.M.C. Foulkes, R.J. Needs, and
G. Rajagopal (unpublished)
\bibitem{cusp}T. Kato,
Communications on Pure and Applied Mathematics {\bf 10}, 151 (1957);
R.~T. Pack, and W.~B. Brown J. Chem. Phys. {\bf 45}, 556 (1966).          
\bibitem{Ceperley}
D. Ceperley, Phys. Rev. B {\bf 18}, 3126 (1978).
\bibitem{Pines-Nozieres} P. Nozieres and D. Pines, \emph{The theory of quantum
 liquids},
(Benjamin, New York, 1966). 
\end{thebibliography}
\end{document}